\documentclass[twocolumn,showpacs,amsmath,amssymb,prb,footinbib,floatfix]{revtex4}

\usepackage{graphicx}% Include figure files
\usepackage{dcolumn}% Align table columns on decimal point
\usepackage{bm}% bold math

\usepackage{helvet}
\usepackage{amssymb}
\usepackage{amsmath}
\usepackage{amsfonts}
\usepackage{amssymb}
\usepackage{hyperref}
\usepackage{color}

\usepackage{xcolor}
\definecolor{red}{rgb}{0.7,0,0}%darkred MIT
\definecolor{green}{rgb}{0.,0.35,0.}%darkgreen
\definecolor{blue}{rgb}{0.2,0.2,0.7} %beamer@blendedblue
\definecolor{black}{rgb}{0.15,0.15,.15}%not too black

\makeindex

\begin{document}

\title{Estimating Quasi-long-range Order via R\'enyi Entropies}

\date{\today}

\author{ M. Dalmonte}
\affiliation{Dipartimento di Fisica dell'Universit\`a di Bologna and INFN, via Irnerio 46, 40126 Bologna, Italy}

\author{E. Ercolessi}
\affiliation{Dipartimento di Fisica dell'Universit\`a di Bologna and INFN, via Irnerio 46, 40126 Bologna, Italy}

\author{L. Taddia}
\affiliation{Dipartimento di Fisica dell'Universit\`a di Bologna and INFN, via Irnerio 46, 40126 Bologna, Italy}

\begin{abstract}
We show how entanglement entropies allow for the estimation of quasi-long-range order in one dimensional systems whose low-energy physics is well captured by the Tomonaga-Luttinger liquid universality class. First, we check our procedure in the exactly solvable XXZ spin-$1/2$ chain in its entire critical region, finding very good agreement with Bethe ansatz results. Then, we show how phase transitions between different dominant orders may be efficiently estimated by considering the superfluid-charge density wave transition in a system of dipolar bosons. Finally, we discuss the application of this method to multispecies systems such as the one dimensional Hubbard model. Our work represent the first proof of a direct relationship between the Luttinger parameter and R\'enyi entropies in both bosonics and fermionic lattice models.
\end{abstract}

\pacs{05.70.Jk, 03.67.Mn, 71.10.Pm, 75.10.Pq}

%apertis verbis, rosier

%\pacs{ 05.30.Jp, 71.10.Pm, 03.75.Lm}

%03.75.Lm===Tunneling, Josephson effect, Bose-Einstein condensates in periodic potentials, solitons, vortices, and topological excitations (see also 74.50.+r Tunneling phenomena; Josephson effects in superconductivity)

%05.30.Jp===Boson systems (for static and dynamic properties of Bose-Einstein condensates, see 03.75.Hh and 03.75.Kk; see also 67.10.Ba Boson degeneracy in quantum fluids)

%71.10.Pm===Fermions in reduced dimensions (anyons, composite fermions, Luttinger liquid, etc.) (for anyon mechanism in superconductors, see 74.20.Mn)

%37.10.Jk===Atoms in optical lattices 

%05.70.Jk 	Critical point phenomena (for quantum critical phenomena in superconductivity, see 74.40.Kb)

%03.67.Mn 	Entanglement measures, witnesses, and other characterizations 

%03.67.Ac 	Quantum algorithms, protocols, and simulations 

%75.10.Pq 	Spin chain models 

%64.70.Tg 	Quantum phase transitions (for quantum Hall effects aspects, see 73.43.Nq in electronic structure of surfaces, interfaces, thin films, and low dimensional structures)

\maketitle
 
\section{Introduction}  Spontaneous symmetry breaking (SSB) has been a central concept in theoretical physics in the last decades, with applications ranging from high-energy to condensed matter systems. However, since the seminal contributions by Mermin, Wagner and Hohenberg\cite{mermin1966}, it has been clear that phase transitions (PT) \cite{sachdev_book} in low dimensional systems cannot in general be described in the context of SSB due to the lack of finite order parameters associated with the breaking of continuous symmetries such as, e.g, translational invariance. In one dimension (1D), gapless systems are usually characterized by {\it dominant orders}, embodied in the asymptotic algebraic decay of correlation functions which determine response functions; such situation is usually referred to as quasi-long-range order (QLRO)\cite{bosonization,bosonization2}, in contrast to true-long-range order associated with SSB. Computing exact correlation functions remains a very challenging task even for exactly solvable models such as the XXZ spin-$1/2$ chain \cite{bosonization}, but the possibility to describe low-energy properties of 1D systems in terms of conformal field theories (CFT)\cite{difrancesco_book} has had a notable impact on the characterization of PTs thanks in particular to the analogy between the compactified boson theory and Tomonaga-Luttinger Liquids (TLL)\cite{bosonization,bosonization2,haldane1981} described by the following Hamiltonian:
\begin{equation}\label{H_TLL}
\mathcal{H}=(v/2\pi)\int \; dx \left[(\partial_x \vartheta)^2/K+K(\partial_x\varphi)^2\right].
\end{equation}
Here, $v$ is the sound velocity, $\vartheta,\varphi$ are conjugated density and phase bosonic fields and $K$ is the TLL parameter, related to the compactification radii  $R_\vartheta= 1/R_\varphi$ of the fields via $K=1/(4\pi R_\vartheta^2)$\cite{bosonization}. TLLs have attracted increasing interest in recent times due to a large number of physical systems whose microscopic description is well approximated by Eq. \ref{H_TLL}: signatures of TLL physics has been predicted and observed in a series of setups ranging from typical condensed matter ones such as carbon nanotubes\cite{bockrath1999} to highly tunable ultracold atomic gases\cite{kinoshita2004}.

In a TLL, the asymptotic decay of correlation functions is entirely encoded in $K$, which becomes the relevant quantity when determining QLRO. For exactly solvable models, $K$ may be estimated through Bethe Ansatz (BA)\cite{bosonization2}; however, such a technique is not in general available, and one has to resort to perturbative or numerical methods. 

Moreover, even thought simulation techniques for 1D systems have become extremely precise in recent years, reliable estimates of $K$ for complex systems such as multispecies Hubbard-like models, describing bosonic and fermionic mixtures\cite{capponi2008}, and frustrated Heisenberg chains \cite{sudan2009} are still challenging. The aim of this paper is to show that QLRO in 1D systems can be accurately determined from only ground state entanglement properties of a microscopic lattice model whose low-energy physics is encoded into the TLL universality class by computing the so called R\'enyi entropies (REs); in addition, we will show how PTs between different dominant orders may be shaped by calculating REs employing the density-matrix renormalization group (DMRG) algorithm\cite{white1992}. The general analysis method is presented in Sec. \ref{meth}, and applications to Heisenberg chains, dipolar bosons and the Hubbard model are described in Sec. \ref{chain_ref}, \ref{dipbos_sec} and \ref{hm_sec} respectively. Finally, Sec. \ref{concl_sec} contains a summary of all results. 

\section{Method}\label{meth}  We will consider a bipartite system $A\cup B$ of size $L$,  where $A$ is an interval of length $l$, and the corresponding REs  defined as  
\begin{equation}
S_{\alpha}(l)=(1-\alpha)^{-1}\log_2{\rm Tr}_B\rho_{l}^{\alpha}
\end{equation}
 where $\rho_{l}$ is the reduced density matrix obtained by having traced out the $A$'s degrees of freeedom. For a CFT, it is known that\cite{holzhey1994,calabrese2004}:
\begin{equation}\label{S_cft}
S_{\alpha}(l)=\frac{c(1+1/\alpha)}{6\eta}\log_2 \left(\frac{\eta L}{\pi}\sin(\pi l/L)\right) + c_{\alpha}'
\end{equation}
where $c$ is the central charge ($c=1$ for a TLL), $\eta=1,2$ for periodic/open boundary conditions (PBC/OBC) and $c_{\alpha}'$ is a model dependent constant not depending on $l$. In a TLL, except for the von Neumann entropy  ($\alpha=1$) under PBC, all REs with $\alpha>1$ are subject to subleading corrections of the form\cite{calabrese2010,cardy2010}:
\begin{equation}\label{S_oscil}
S_{\alpha}^{osc}(l)= F_{\alpha}(l/L)\cos(2k_Fl+\omega)\left|\frac{2\eta L}{\pi}\sin\frac{\pi l}{L}\sin(k_F)\right|^{-p_{\alpha}}
\end{equation}
where $k_F$ is the Fermi momentum,  $F_{\alpha}$ is a universal scaling function of $l/L$ and $\omega$ is an $l$-independent phase shift. The information about QLRO is encoded into the decay exponents $p_{\alpha}=2K/(\eta\alpha)$. Such oscillating corrections have been confirmed for a series of models both analytically and numerically \cite{laflorencie2006,calabrese2010,fagotti2011,xavier2011}, and their relationship with correlations functions have been deepened in Ref.\cite{song2010}. Parity effects in REs thus provide a useful tool to determine the TLL parameter exclusively from ground state properties. There might be, however, notable difficulties. Indeed, the amplitude of the oscillations may be  orders of magnitude smaller than the CFT contributions, thus making quantitative estimates challenging\cite{xavier2011}. Furthermore, an even more serious problem is given by the scaling function $F_{\alpha}$: its explicit form is {\it a priori} not known and the assumption that it is independent of $l/L$ is not in general justified. Nevertheless, we argue that $K$ can be accurately extracted from the following quantity
\begin{eqnarray}\label{dS}
dS_{\alpha}(L)&\equiv& S_{\alpha}(L/2)-S_{\alpha}(L/2-\pi/(2k_F))=\\
&=& a_1/L^2+ \cos(k_FL+\omega)L^{-\frac{2K}{\eta\alpha}}(a_2+\mathcal{O}(1/L))\nonumber
\end{eqnarray}
by studying its dependence on $L\gg 1$. Here, $a_1$ and $a_2$ are constants and finite size corrections to $c$ and $K$ are given by higher order contributions which are negligible for large system sizes (see Appendix \ref{app_a} for further details on the fitting procedure). The advantages of this method rest on the fact that this quantity {\it i)} does not require any {\it a priori} knowledge of $F_{\alpha}(l/L)$, {\it ii)} can be easily extracted from DMRG procedure for arbitrary $\alpha$, and {\it iii)} may be evaluated for different $\alpha$'s through a single simulation, thus allowing to consider the RE with better scaling properties, such as comparable oscillation magnitude with respect to the CFT contribution.

\begin{figure}[t]{
\begin{center}
\includegraphics[width=4.23cm]{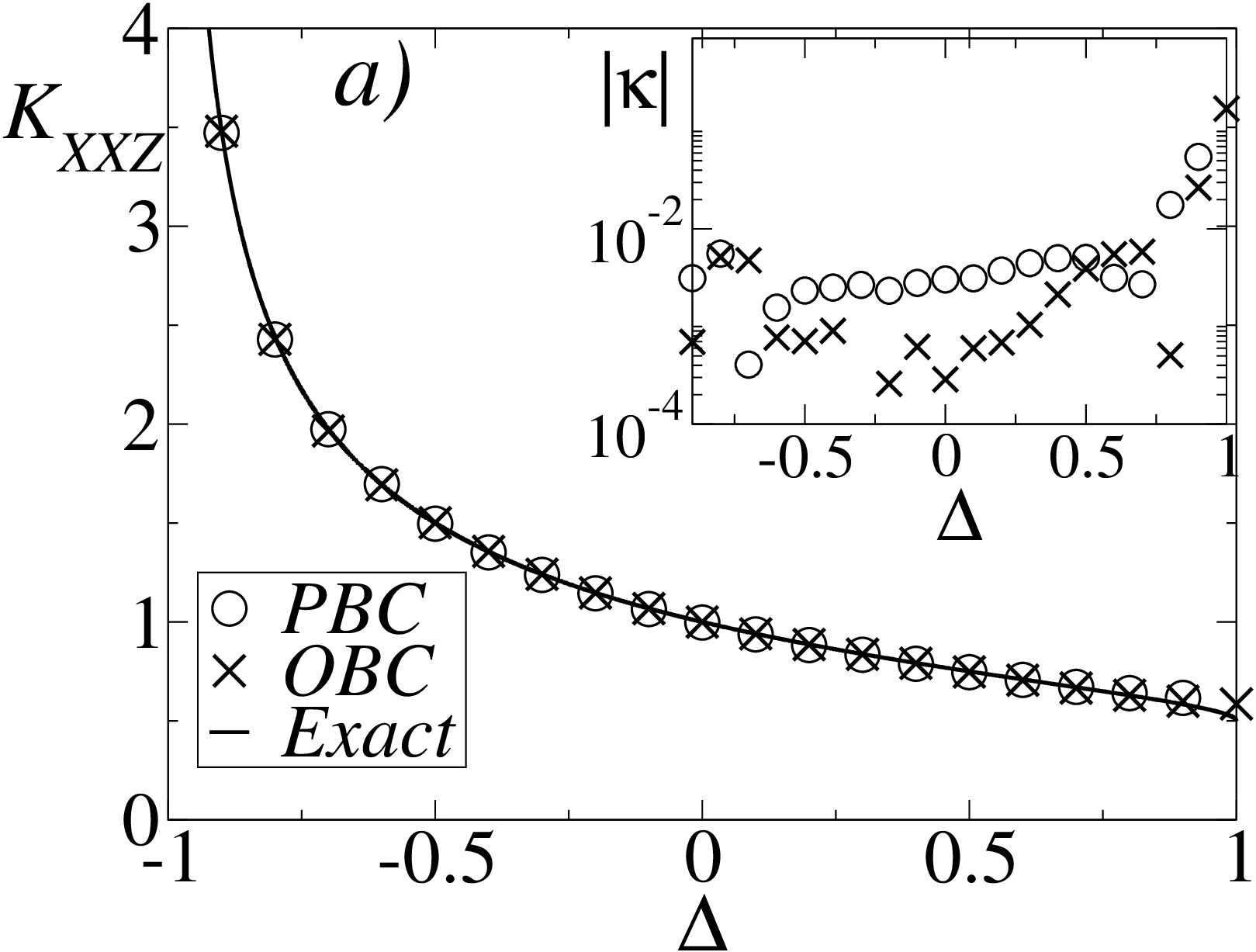}
\includegraphics[width=4.23cm]{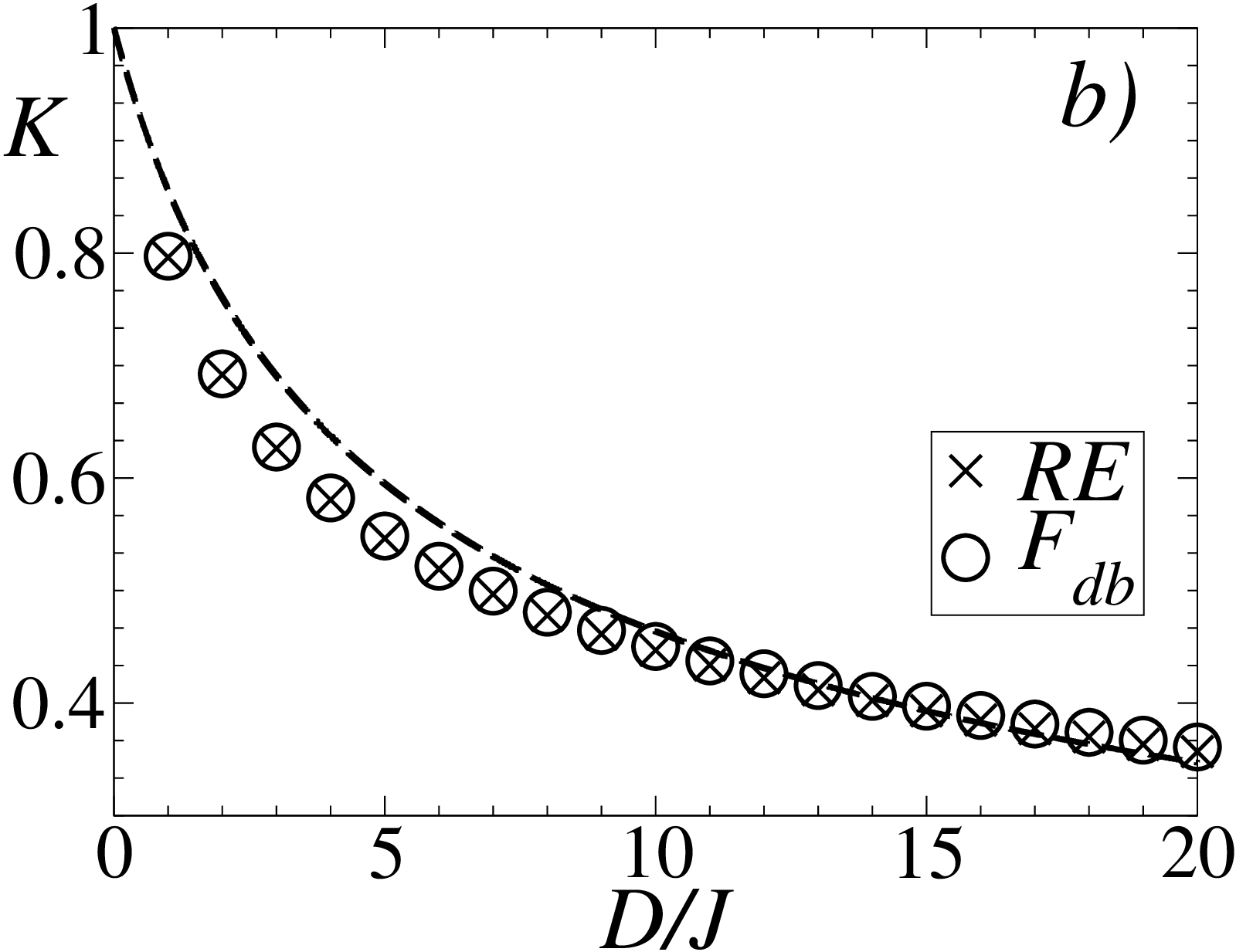}
\caption{Estimate of the TLL parameter from numerical calculations. Panel {\it a}:  $K$ extracted from numerical data on REs (circles and crosses)  and exact one,  $K_{BA}$,  from  BA solution (solid curve) for a spin-$1/2$ XXZ chain. Inset: relative difference $\kappa=K/K_{BA}-1$ as a function of $\Delta$ in linear-log scale. Panel {\it b}: numerical results for hard-core dipolar bosons at $\bar{n}=1/4$; estimates of $K$ based on fluctuations ($F_{db}$) and REs under OBC are in very good agreement in a broad parameter range. The dashed line is an analytical result from Ref.\cite{DPZ} in the continuum limit.}
 \label{fig_comp}
 \end{center}
 }
\end{figure}

In the following, we will prove how $dS_{\alpha}$ can provide quantitative information about QLRO in a series of 1D models described by a TLL. We will first consider a spin-$1/2$ XXZ chain, and compare RE results with those obtained with the BA, finding excellent agreement in the entire critical regime except close to the antiferromagnetic point, as summarized in Fig. \ref{fig_comp}a.  Then, we will investigate the superfluid/charge-density-wave (SF/CDW) transition in a bosonic gas with dipolar interactions, and will compare our results with those gotten from other independent methods, as shown in Fig. \ref{fig_comp}b and more technically in Appendix \ref{app_b}. Finally, the 1D attractive Hubbard model would be considered as a paradigmatic case for estimating the TLL parameters in multicomponent systems.

\section{Spin-$1/2$ XXZ chain}\label{chain_ref} 
\begin{figure}[t]{
\begin{center}
\includegraphics[width=4.26cm]{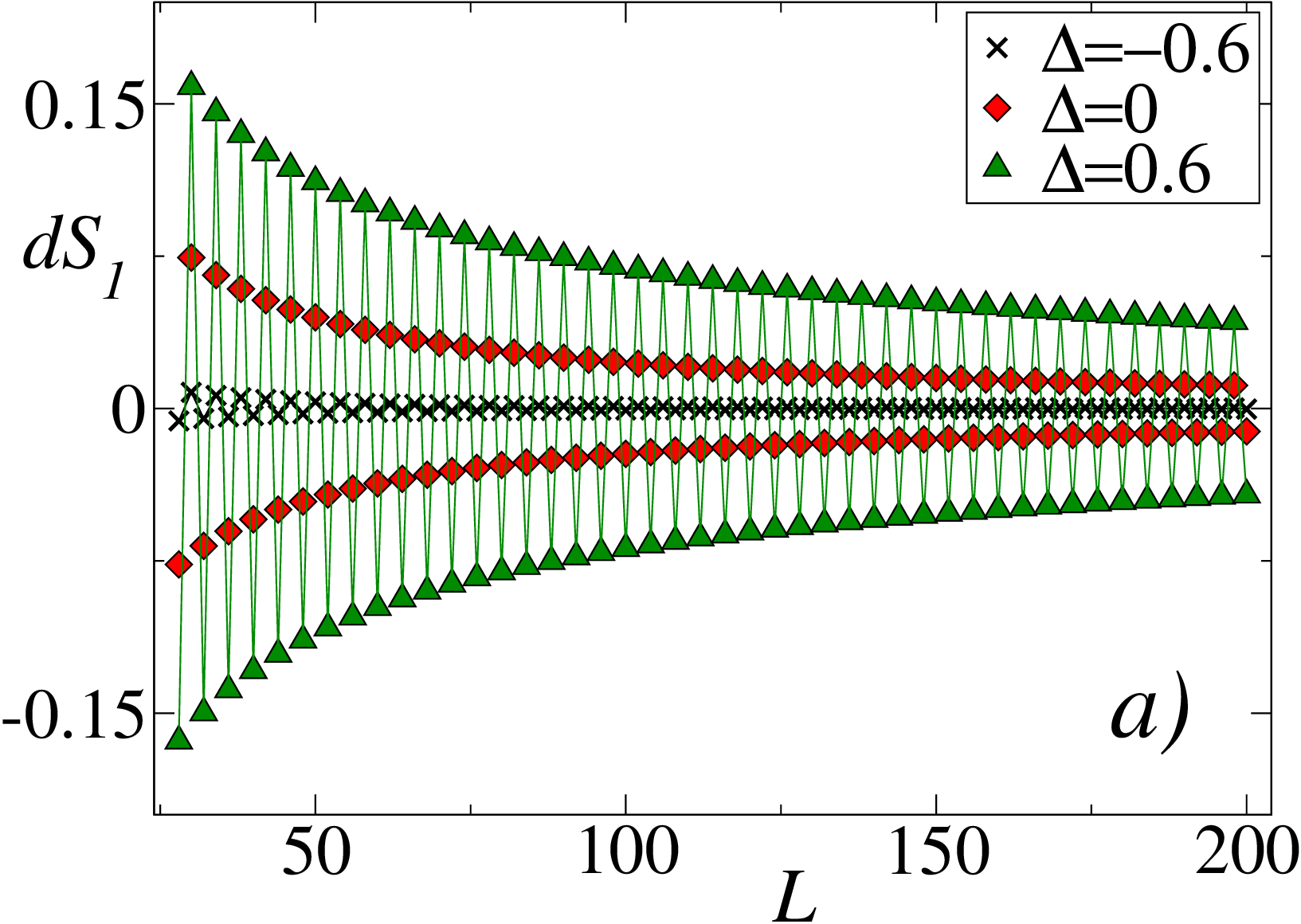}
\vspace{1.4mm}
\includegraphics[width=4.26cm]{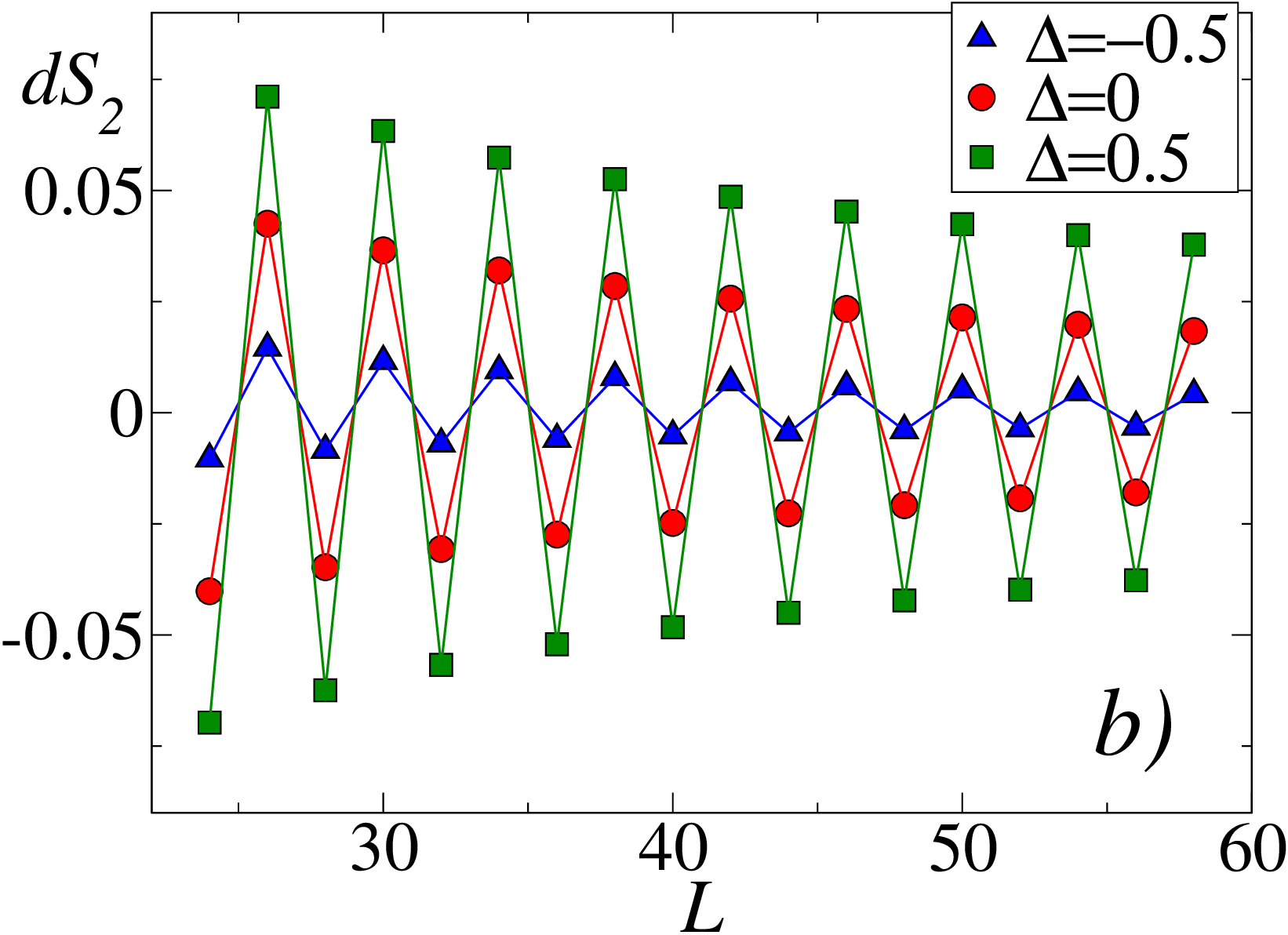}
\includegraphics[width=4.26cm]{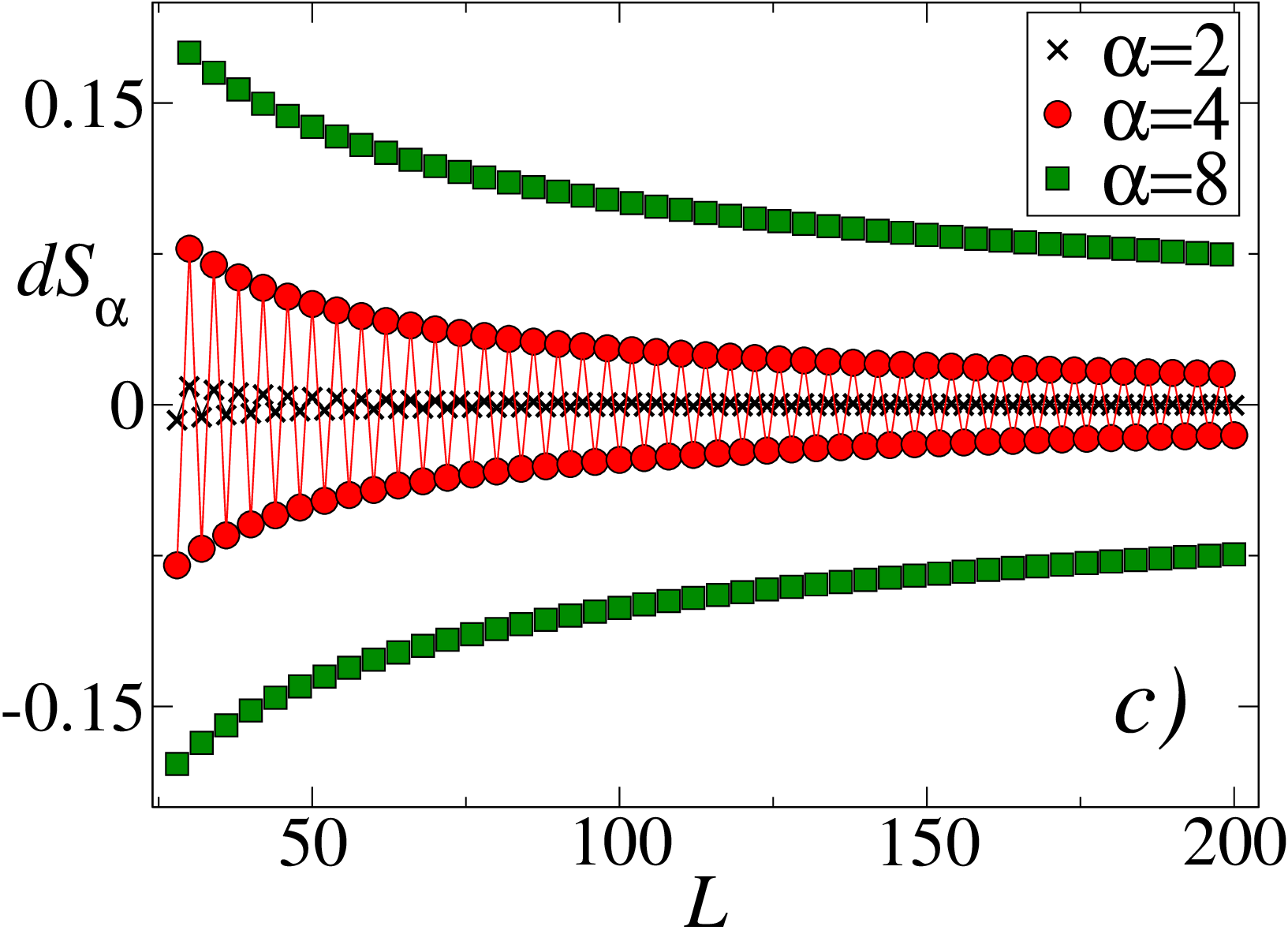}
\includegraphics[width=4.26cm]{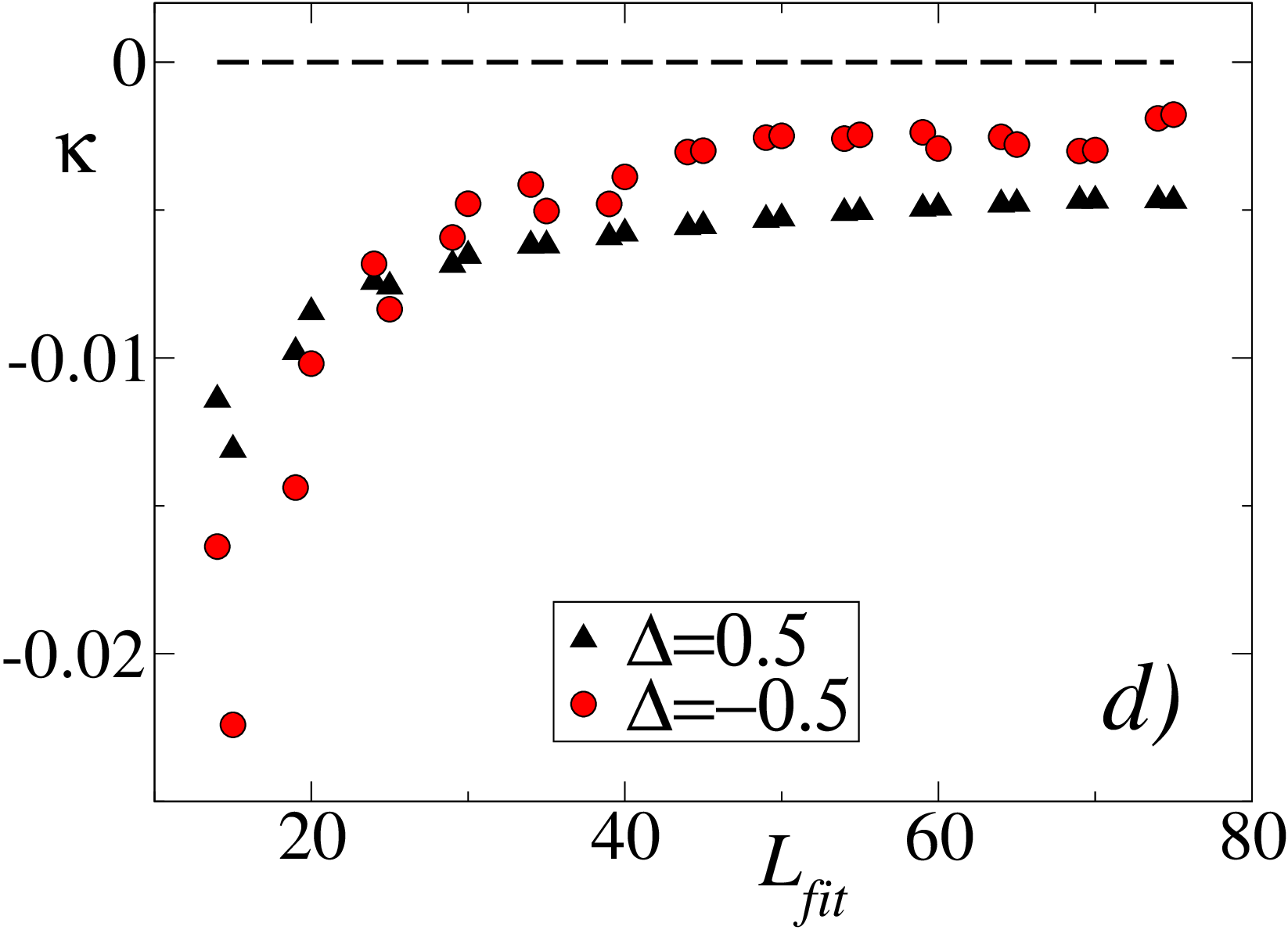}
\caption{(color online). Numerical analysis of the XXZ model. Panels {\it a,b}: $dS_1$ (OBC) and $dS_2$ (PBC) as a function of $L$ for various $\Delta$; the magnitude of oscillations decreases with $\Delta$ in both cases. Panel {\it c}: $dS_{\alpha}$ (OBC) at $\Delta=-0.9$ for various $\alpha$ as a function of $L$; the amplitude of the oscillations is very small for $\alpha<2$; lines are typical best fits. Panel {\it d}: discrepancy between exact and numerical value of $K$, $\kappa=K/K_{BA}-1$, as a function of the fit points $L_{fit}$. Here, the fitting procedure takes into account  system sizes in the interval $[52, 52+2L_{fit}]$ with OBC: the relative error is under $1\%$ even for $L\lesssim 100$.  }
 \label{fig_XXZ}
 \end{center}
 }
\end{figure}

As a first case of interest, we consider the spin-$1/2$ Heisenberg chain:
\begin{equation}\label{H_XXZ}
H_{XXZ}=-\sum_{i=1}^L(S_i^xS_{i+1}^x+S_i^yS_{i+1}^y-\Delta S_i^zS_{i+1}^z)
\end{equation}
in its critical region $-1<\Delta\leq1$, where it is described by a TLL theory:  the TLL parameter can be extracted from BA, $K_{BA}=\pi/2\arccos(-\Delta)$, and $k_F=\pi/2$. Recent numerical results\cite{calabrese2010,xavier2011} have shown that 
Eq. \ref{S_oscil} holds for both  periodic (PBC) and open boundary
conditions (OBC) for several values of $\Delta$: this system can then provide a very accurate check for the
proposed procedure. 
We have investigated Eq. \ref{H_XXZ} in the entire critical region, with both OBC and PBC, by employing numerical
simulations based on the DMRG algorithm, computing several REs ($0.5\leq\alpha\leq 100$) 
with systems up to L=200/60 (OBC/PBC); in order to provide accurate estimates of all REs, we have 
applied several sweeps at each system size during the infinite-size procedure, and considered a number
of states such that the truncation error of the last step is usually smaller than $10^{-9}/10^{-8}$ (OBC/PBC).
As already known, we observe that oscillations are usually much more pronounced 
for large values of $\alpha$; in addition, for OBC, they are present also for $\alpha<1$, following the same periodic behavior of all other REs. For $\Delta\simeq0$, a raw fit
of the RE at $L=200$ usually gives already a good estimate of $K$, albeit the relative small modulus of the oscillations
induces large errors for small $\alpha$'s.  Computing $d S_{\alpha}$, instead, turns out to be a very efficient way to estimate $K$. In Fig. \ref{fig_XXZ}a-c,
we plot $dS_{\alpha}$ for several values of $\alpha$ and $\Delta$ for both OBC and PBC: all curves 
present oscillations whose magnitude increases with $\alpha$. Best fits of Eq. \ref{dS}
agree very well with the numerical datas except for small systems sizes ($L\leq24$), which are therefore rejected. The addition of corrections that include finite size effects on $K$ or $c$ does not alter the fitted value of $K$. 
The first main result of this work is plotted in Fig.\ref{fig_comp}a: the values of $K$ extracted through $dS_{\alpha}$ for PBC and OBC are plotted against the exact BA values in the entire critical region $-1<\Delta\leq 1$. Estimates of $K$ with OBC/PBC are taken at $\alpha=1,2$ respectively, except for $\Delta\leq -0.8$, where oscillations are so small that a good fitting procedure requires larger values of $\alpha$ ($\alpha=4/8$). As it can be seen from the inset, the intrinsic error of this procedure is  usually of order $10^{-3}$, except close to the Berezinskij-Kosterlitz-Thouless (BKT) transition, where marginal operators induce quantitatively relevant corrections to $dS_{\alpha}$\cite{calabrese2010}. Deviations from the exact value as function of the system size are plotted in Fig. \ref{fig_XXZ}d, which shows that relatively small systems of $L=80$ sites already achieve good accuracies, $\kappa\lesssim 0.01$, and that strong oscillations in  the estimate of $K$ with respect to the number of points employed in the fit signal strong finite-size effects, which are strongly reduced considering larger systems. We can thus conclude that, away from BKT transitions where possible logarithmic corrections can emerge, $dS_{\alpha}$ provides a very accurate estimate of the TLL parameter in the XXZ model.

\section{Dipolar bosons in a single tube}\label{dipbos_sec}  
\begin{figure}[t]{
\begin{center}
\includegraphics[width=4.26cm]{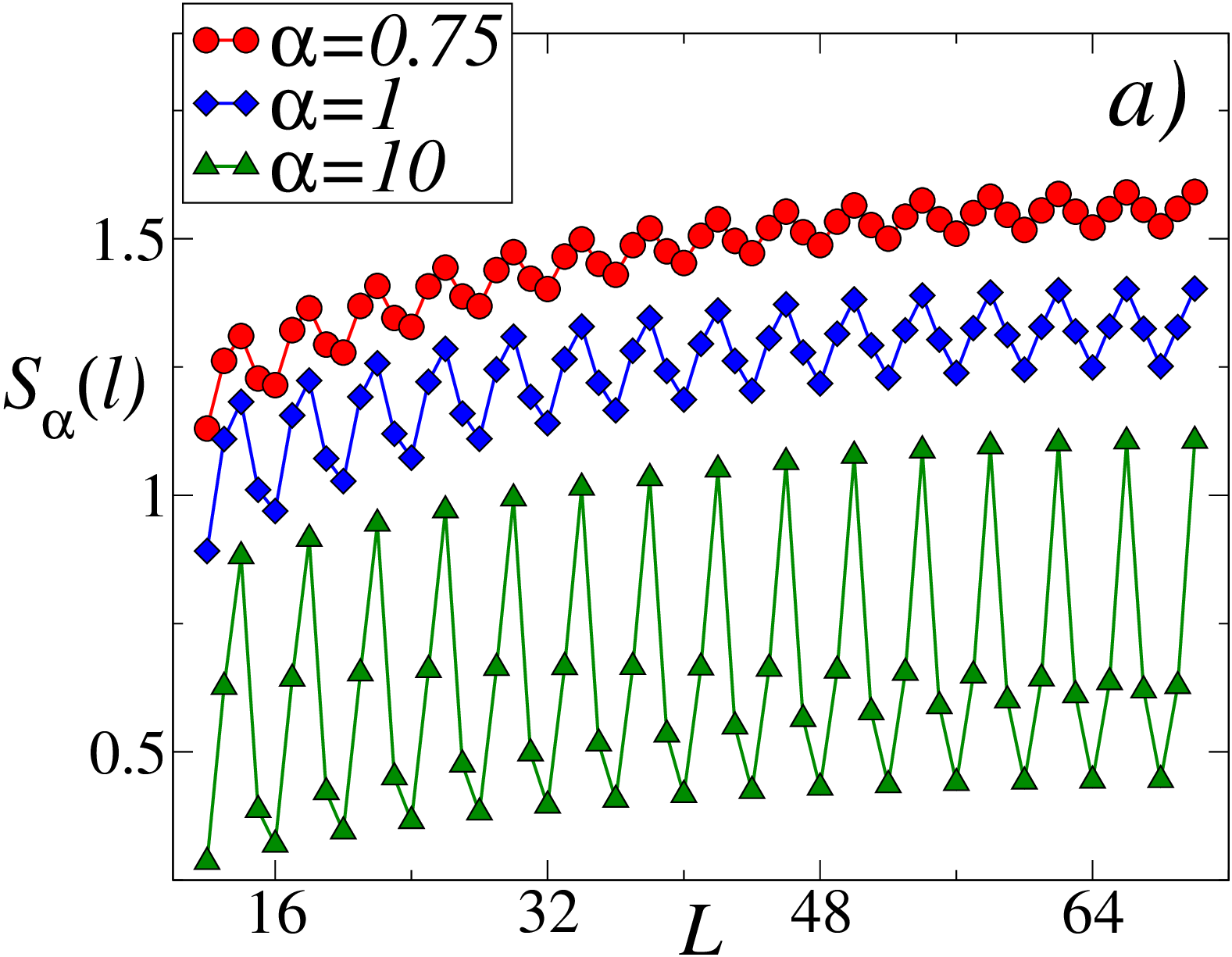}
\vspace{1.4mm}
\includegraphics[width=4.26cm]{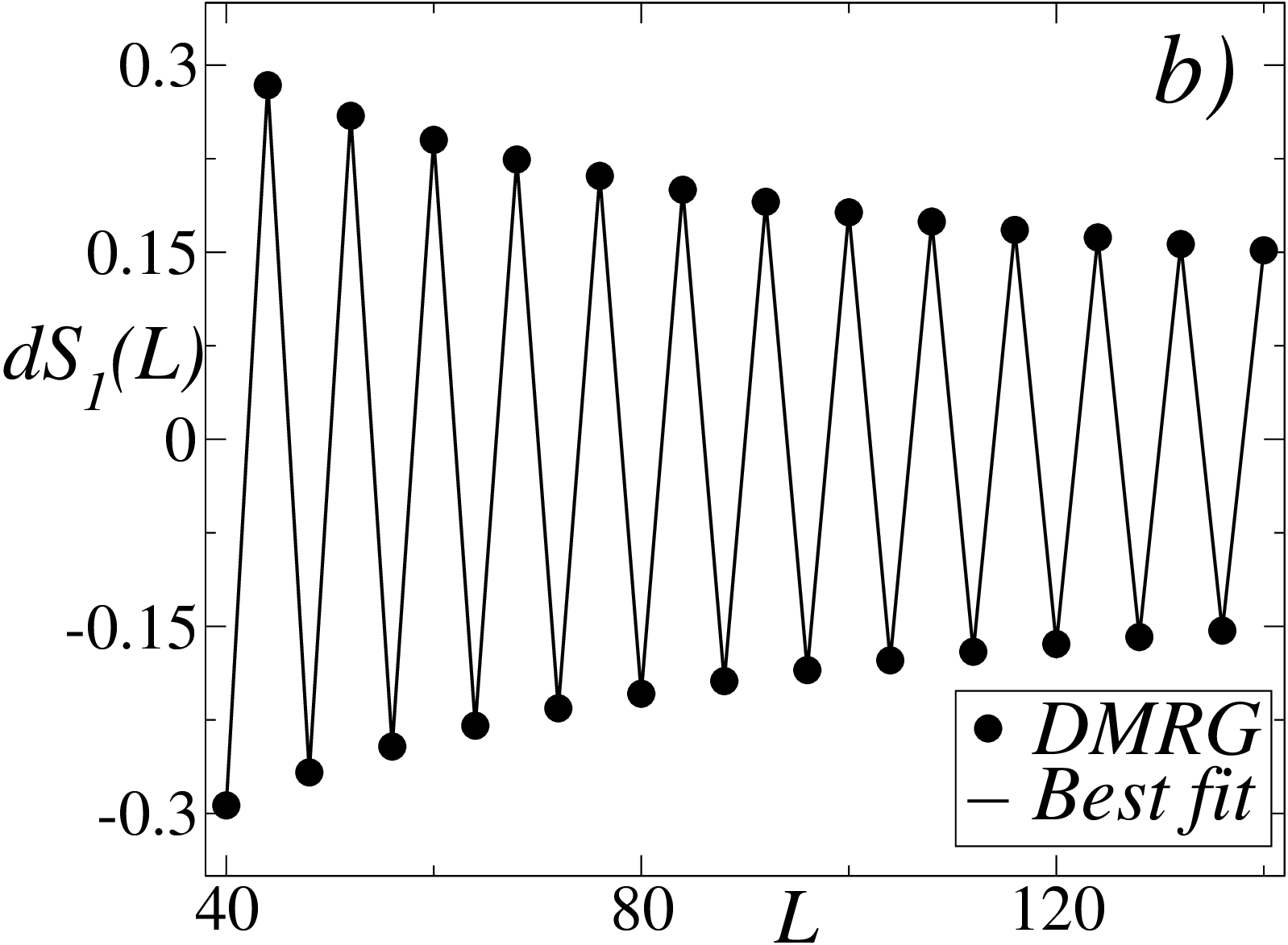}
\includegraphics[width=4.26cm]{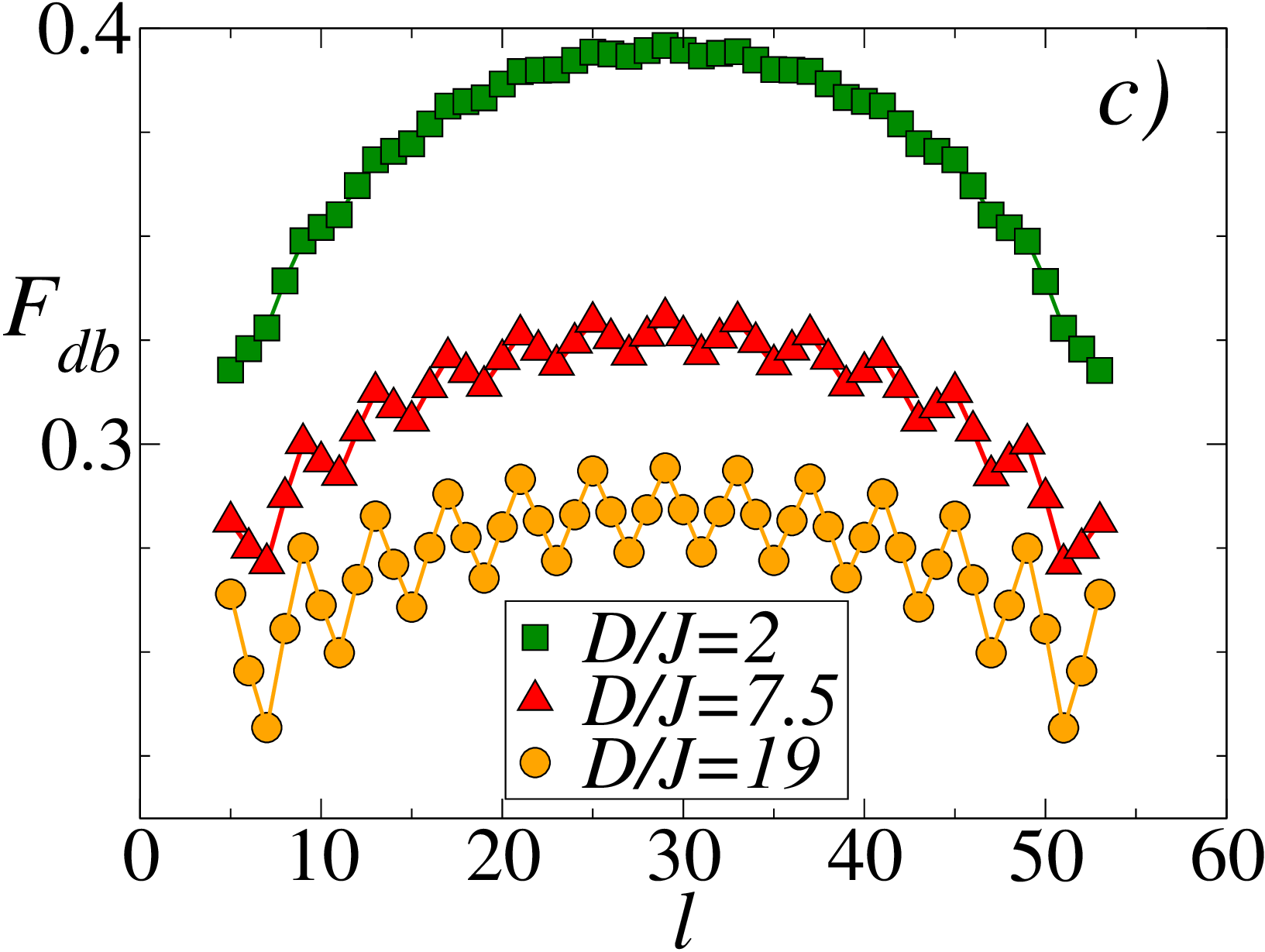}
\includegraphics[width=4.26cm]{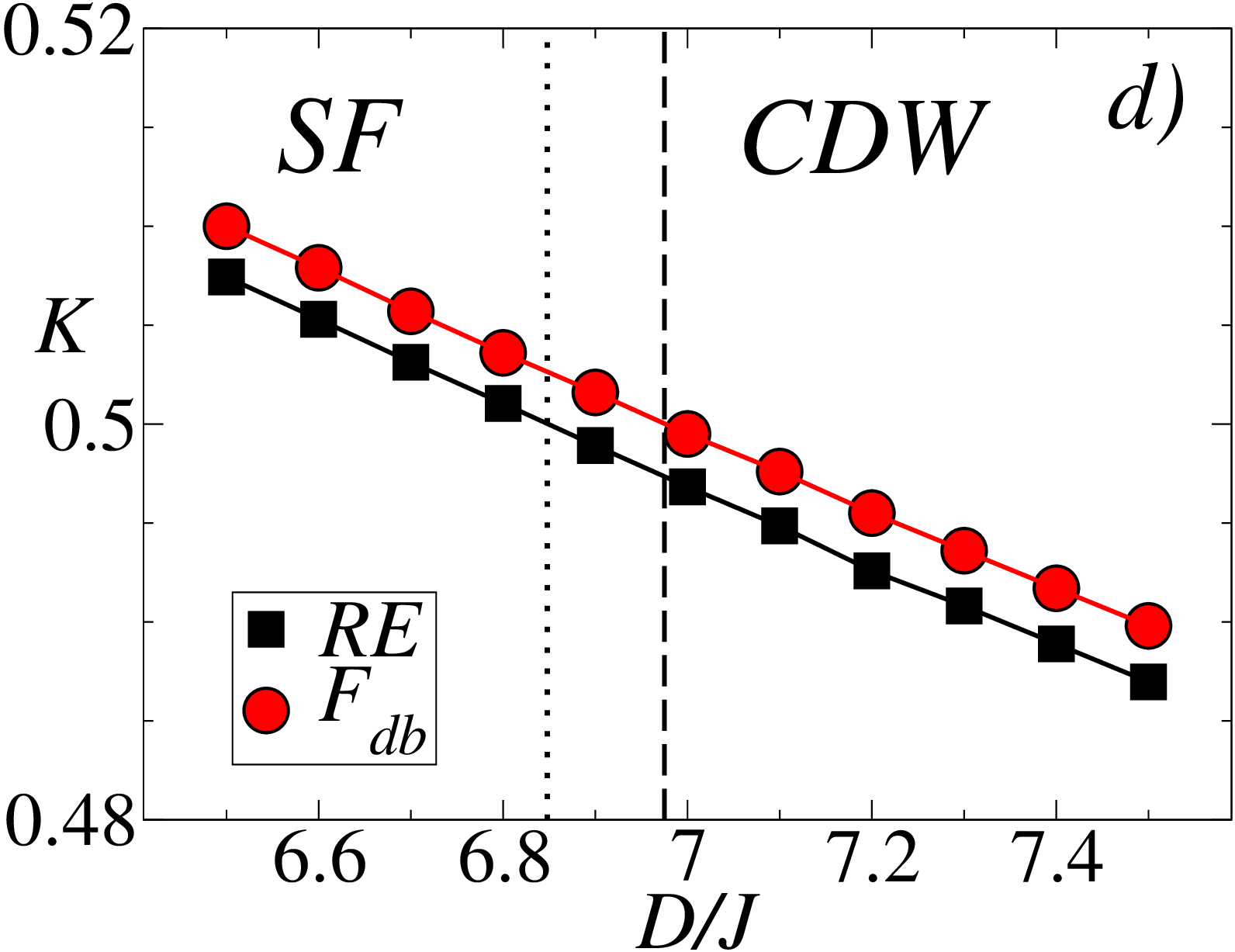}
\caption{(color online). Estimate of the TLL parameter for dipolar bosons. Panel {\it a}: typical scaling of $S_{\alpha}$ at $D/J=5$ and, from top to bottom,  $\alpha=0.75, 1,  10$ in a $L=140$ site system with OBC. Oscillations are present for every $\alpha$, their magnitude increasing with $\alpha$. Panel {\it b}: scaling of $dS_1$ as a function of $L$ for $D/J=5$ with OBC: full circles denote numerical data, black line is the best fit of Eq. \ref{dS}, from which we extract $K_{db}=0.5436$. Panel {\it c}: $F_{db}(l)$ for different values of $D/J$; lines are best fits (see text).  Panel {\it d}: SF-CDW phase transition as estimated with both $F_{db}$ (dotted) and RE (dashed line) with OBC; solid lines are interpolations.}
 \label{fig_S_dipbos}
 \end{center}
 }
\end{figure}

In 1D systems, PTs between phases with different dominant order can appear even within the same gapless region,
that is, a TLL may exhibit different phases as a function of $K$\cite{bosonization,bosonization2}. A typical example is given by bosonic particles interacting
through a non-local repulsive potential, as realized in one dimensional tubes of polar molecules or magnetic atoms with
dipole moment aligned perpendicularly to the tube via a dc electric field\cite{LewensteinReview,citro_roscilde,DPZ}. When loaded onto an optical lattice\cite{bloch2008}, their effective Hamiltonian is:
\begin{equation}\label{hdb}
H_{db}=-J \sum_{i=1}^L (b^{\dagger}_ib_{i+1}+h.c.)+D \sum_{i<j}\frac{n_in_j}{|i-j|^3}
\end{equation}
where $b^{\dagger}_i,b_i $ are hard-core bosonic creation/annihilation operators on the site $i$, $n_i=b^{\dagger}_ib_i$ and the ratio $D/J$ can be tuned, e.g., by varying the depth of the optical potential. In this case, increasing the interparticle 
dipolar repulsion $D/J$ induces a transition from a SF order with dominant single-particle correlations 
\begin{equation}
B(x)\simeq\langle b_i^{\dagger}b_{i+x}\rangle \simeq x^{-1/2K}
\end{equation}
 to a CDW order with dominant density correlations 
 \begin{equation}
 \mathcal{D}(i,i+x)=\langle n_i n_{i+x} \rangle_c\simeq\frac{K}{2\pi x^{2}}+\cos(2k_F x)x^{-2K}.
 \end{equation}
  This transition occurs at a precise value of the TLL parameter, $K=1/2$\cite{bosonization,bosonization2,DPZ}. We have thus investigated the SF-CDW phase transition at filling $\bar{n}=1/4$\cite{note1} so that $k_F=\pi/4$, truncating the dipolar interaction up to fifth-nearest-neighbors\cite{note2}. Typical results for REs and $dS_{\alpha}$ are plotted in Fig. \ref{fig_S_dipbos}a-b: systems with up to $L=140/60$  have been considered for OBC/PBC respectively, keeping up to $512/1024$ states and employing finite-size sweeps at each even intermediate size.  

\begin{figure}[t]{
\begin{center}
\includegraphics[width=4.25cm]{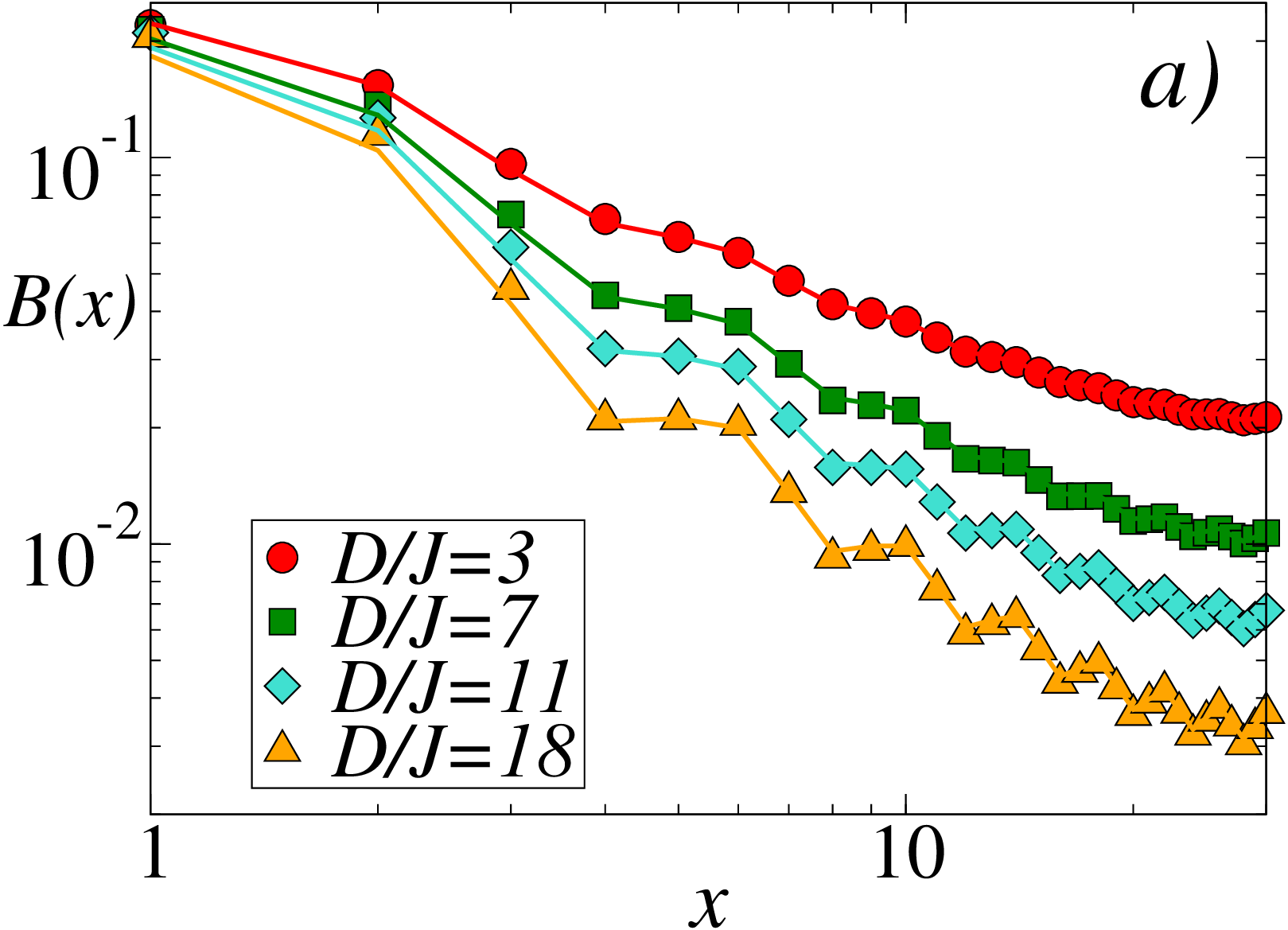}
\includegraphics[width=4.25cm]{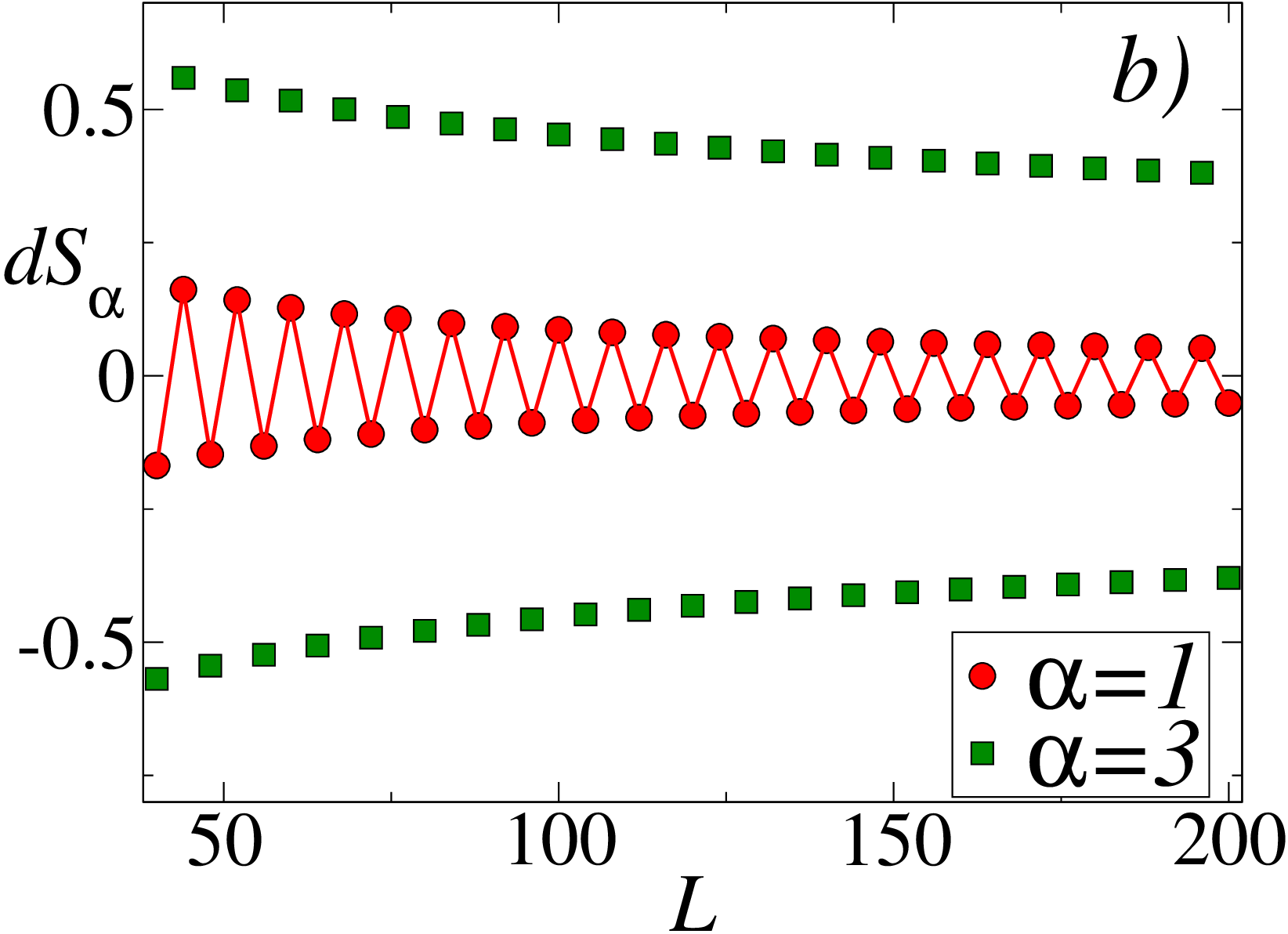}
\caption{(color online). Panel {\it a}: SF correlation $B(x)$ for dipolar bosons; lines are best fits using CFT formulas\cite{cazalilla2004} where we have used $K$ as estimated form  the RE. Panel {\it b}: $dS_{\alpha}$ for the HM at $n=1/4,U=-21$ with OBC; the thin line denotes a best fit with Eq. \ref{dS}  from which $K=1.524$, in good agreement with the BA value $K_{BA}^{(HM)}=1.526$.}
 \label{fig_fdb}
 \end{center}
 }
\end{figure}

Since Eq. \ref{hdb} is not exactly solvable, we have estimated $K$ through other independent methods. In Fig. \ref{fig_comp}b, the values of $K$ extracted form RE are compared with those extracted from density fluctuations\cite{song2010}:
\begin{equation}
 F_{db}(l)=\sum_{i,j<l}\mathcal{D}(i,j)\simeq (K\ln l)/\pi^2.
 \end{equation}
  Typical behaviors of $F_{db}(l)$ are shown in Fig. \ref{fig_S_dipbos}c. The so-obtained estimates of $K$ are in very good agreement in an ample parameter range, as it can be inferred from the data in Fig. \ref{fig_comp}b. 
A snapshot around the SF-CDW PT is given in  Fig. \ref{fig_S_dipbos}d. Still a different method to estimate $K$ is given by level spectroscopy\cite{bosonization2}: it can be checked that also in this case the results  are in good agreement as discussed in Appendix \ref{app_b}. Moreover the estimated $K$ fits very precisely relevant observables such as $B(x)$ (see Fig. \ref{fig_fdb}a). 

%{\color{red}analytical estimate $K_{dp}=(1+1.46\bar{n}\delta/t)^{-1/2}$

\section{1D Hubbard model}\label{hm_sec} TLLs represent the basic element to describe also multicomponent systems such as spin ladders, integer spin chains and multispecies systems. In some fermionic systems, in particular, all but one degrees of freedom may become gapped due to interaction effects, thus making the effective low-energy theory being given by a single TLL. However, the total REs are influenced also by the gapped degrees of freedom, so that Eq. \ref{S_cft} may not allow for a reliable evaluation of the TLL parameter. To investigate this possibility, we have considered, as a paradigmatic example, the 1D Hubbard model (HM)\cite{bosonization2} 
\begin{equation}
H_{HM}=\sum_{i}\big[-\sum_{\sigma=\uparrow,\downarrow}(c_{\sigma,i}^{\dagger}c_{\sigma,i+1}+h.c.)+Un_{i\uparrow}n_{i\downarrow}\big]
\end{equation}
 where $c^{\dagger}_{\sigma, i},c_{\sigma,i}$ are fermionic creation/annihilation operators relative to the species $\sigma$. In the balanced $n_1=n_2=n$, attractive $ U<0$ regime, this system displays spin-charge separation with a gapped spin sector and a gapless charge sector, characterized by a TLL parameter $K_c>1$ which can be calculated via BA\cite{bosonization2,shastry}.  We checked the validity of our approach by considering a fixed density $n=1/4,k_F=\pi/4$, different values of the interactions, $U=-15,-18,-21$ and systems with OBC/PBC with up to $L=200/72$ sites. Typical REs for various $\alpha$'s are plotted  in Fig. \ref{fig_fdb}b. The gapped spin sector appears to play little to none effect on $dS_{\alpha}$, whose behavior  is determined by the gapless charge sector through $K_c=2K$\cite{bosonization2}. Indeed, the estimate of the latter from RE is found to be in excellent agreement (up to $1\%$) with the BA values of Ref. \cite{shastry} in all cases. We interpret this positive result as due to the fact that the contribution of gapped degrees of freedom to REs may be incorporated in the value of the constant term, which does not appear in $dS_{\alpha}$.

\section{Conclusions}\label{concl_sec} We have shown how QLRO can be efficiently estimated through R\'enyi entropies of a single block in a series of 1D models via DMRG calculations. We have tested this technique on single and two-component exactly solvable models such as XXZ spin-$1/2$ chain and the attractive HM. Also, we have calculated the TLL parameter $K$ for hard-core bosons interacting via dipolar interactions, finding very good agreement between entropy results and other methods. Our study, along with related investigations based on REs \cite{furukawa2009}, shows how entanglement  entropies provide an accurate tool to determine critical properties and phase transition in TLLs, as well as  the long-distance decay of correlation functions, in a large variety of 1D systems, such as spin chains, fermionic and bosonic Hubbard-like models. Being less sensitive to finite-size effects and linked to a general and recurrent theoretical background, the technique presented here has several potential applications, such as the determination of correlation functions of composite liquids in atomic and molecular multispecies mixtures\cite{bloch2008}, or in more complex spin chains such as half-integer frustrated Heisenberg models\cite{sudan2009}, where other numerical analysis such as level spectroscopy lose their typical efficiency. 

\section*{Acknowledgements}
We thank P. Calabrese, C. Degli Esposti Boschi and S. Evangelisti for discussions, M. Montanari and S. Sinigardi for technical support, and F. Ortolani for help with the DMRG code.

\appendix

\section{Details on the fitting procedure}\label{app_a}

\begin{center}
\begin{figure}[b]{
\includegraphics[width=7.5cm]{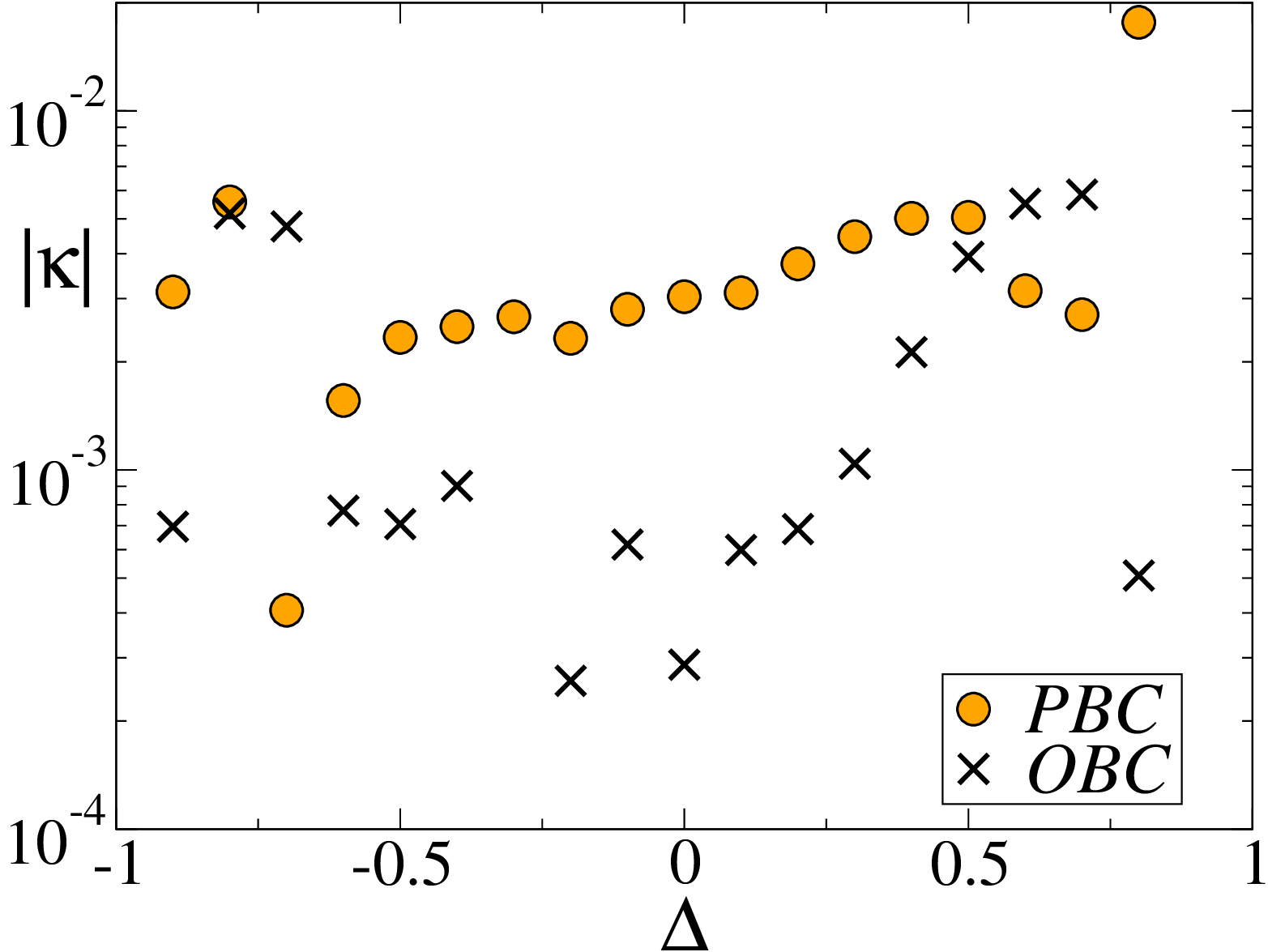}
\caption{(color online). Deviation $\kappa=K/K_{BA}-1$ of the TLL parameter as estimated from RE with respect to the BA results in the XXZ spin-1/2 chain.}
 \label{smfig_XXZ}
}
\end{figure} 
 \end{center}
In this Appendix, we analyze in detail the properties of $dS_{\alpha}(L)$. As noticed in the Sec. \ref{meth}, $dS_{\alpha}$ has two relevant advantages, namely it is not influenced by the non-universal factor $F_{\alpha}(l/L)$, since it depends on $F_{\alpha}(1/2)$ only, and can be straightforwardly calculated via DMRG during the finite-size procedure. The numerical coefficient $a_1$ is obtained from Eq. 3 of the main text:
\begin{equation}
a_1= \frac{\pi^4c(1+1/\alpha)}{48\eta\ln 2k_F^2}
\end{equation}
while $a_2$ is a model dependent constant. Furthermore, additional oscillatory corrections may be considered by expanding the non-universal factor around $L/2$; the first correction, which dominates with respect to the other ones at long-distances, is:
\begin{equation}
a_3\cos(k_FL+\omega)L^{-\frac{2K}{\eta\alpha}-1}.
\end{equation}
Even including these corrections, a best fit of $dS_{\alpha}$ requires a limited number of points, since one is left with only three fitting parameters ($K,a_2,a_3$). Another remarkable fact is that our procedure is weakly affected by finite-size corrections to both $c$ and $K$. Scaling corrections of the form $c(L)=c+\xi_1 L^{-\omega_1},K(L)=K+\xi_2 L^{-\omega_2}$ induce additional terms in $dS_{\alpha}$ whose typical scaling is $L^{-2-\omega_1},(\ln L) L^{-p_{\alpha}-\omega_2}$ respectively. Except for very small values of $\alpha$, these corrections can be safely neglected. We have verified the consistency of this approximation in several cases, for both models considered, finding no appreciable discrepancy between the TLL parameter value extracted with or without terms which include finite-size correction of both $c$ and $K$. 
 
 Another crucial point in our procedure is the choice of the appropriate RE in order to evaluate $K$. Very small values of $\alpha$ can suffer from larger DMRG errors and may have very small absolute values; on the contrary, larger $\alpha$'s usually show too large oscillations, inducing considerable errors in the fitting procedure. The best choice is usually an intermediate value of $\alpha$ that represents a good compromise.

A basic check of the accuracy of $dS_{\alpha}$ in the context of the spin-1/2 XXZ chain has already been discussed in Sec. \ref{chain_ref}; here, we present an enlarged plot of $\kappa$ (Fig. \ref{smfig_XXZ}), which represents the discrepancy of the RE results with respect to the exact Bethe ansatz one. For both PBC and OBC, $\kappa$ is well below 0.01 except close to the BKT transition. Nevertheless, this intrinsic limitation is well under control, being the critical value of $K$ at the BKT point usually known from field theoretical considerations.

\begin{figure}[b]{
\begin{center}
\includegraphics[width=4.25cm]{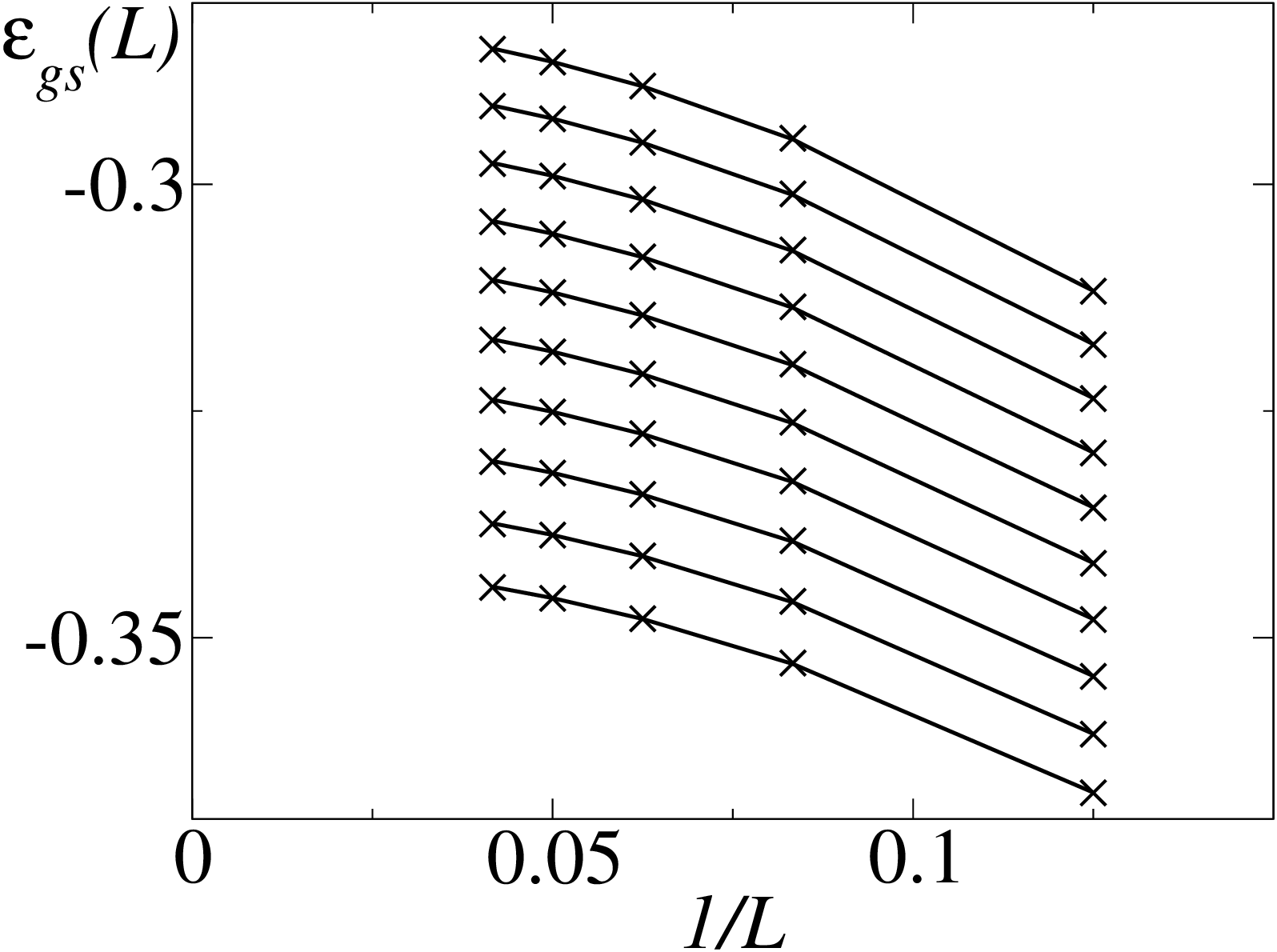}
\includegraphics[width=4.25cm]{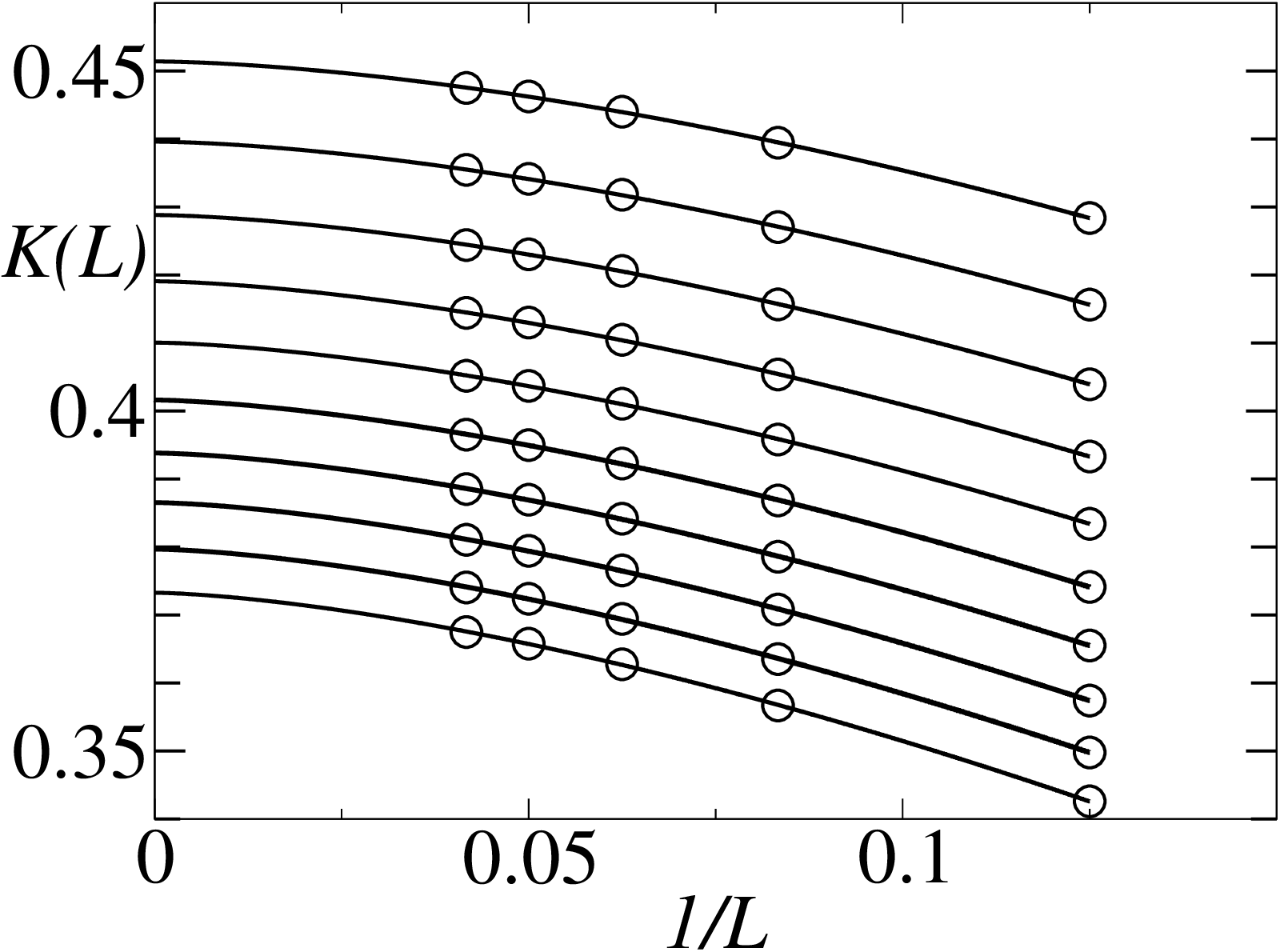}
\caption{Level spectroscopy results from exact diagonalization for a system of dipolar bosons with up to $L=24$ sites. Left panel: $\epsilon_{gs}$ for integer values of $D/J$ from 20 to 11 (top to bottom). Right panel: finite size scaling of $K$ with $D/J$ integer from 11 to 20 (top to bottom). }
 \label{fig_LS}
 \end{center}
 }
\end{figure}

\section{Evaluation of $K$ via density fluctuations and level spettroscopy}\label{app_b}

TLL physics of one-dimensional dipolar bosons has been investigated within a series of methods in the continuum limit, that is, in absence of a periodic potential\cite{citro_roscilde,DPZ}; however, no quantitative prediction in a lattice is currently available, so we had provided a series of additional estimates of $K$ as a reliability check of the RE results. The first quantity of interest are density fluctuations of the kind:
\begin{equation}\label{Fdb}
F_{db}(l)=\sum_{j,k=1}^{l}[\langle n_{j}n_k\rangle-\langle n_j\rangle\langle n_k\rangle]=\frac{K\ln l}{\pi^2}+A_1+\mathcal{O}(l^{2K})\nonumber
\end{equation}
which have been shown to provide a very good estimate of $K$ in the XXZ model\cite{song2010}. We calculated $F_{db}$ for systems with PBC and up to $L=60$ sites for all points considered in the article: these results are in very good agreement with the RE estimates in the entire parameter range, as shown in Fig. \ref{fig_comp}b.

\begin{figure}[t]{
\begin{center}
\includegraphics[width=6.5cm]{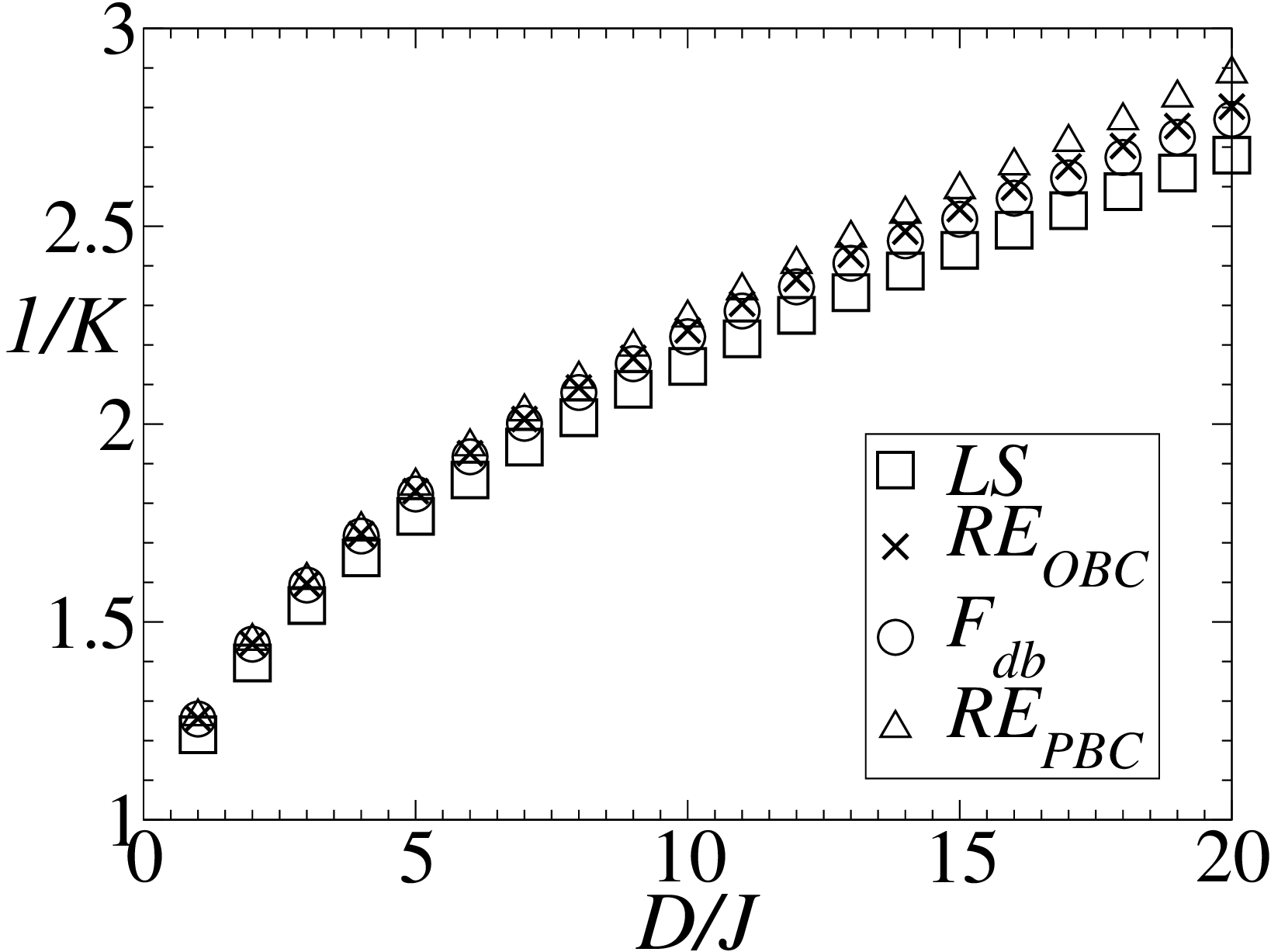}
\caption{Inverse TLL parameter for dipolar bosons for $D/J\in [1,20]$. REs under OBC and fluctuations, which are less affected by finite-size effects, are in very good accordance in the entire parameter regime; LS results overestimate systematically $K$, whereas REs with PBC have a $\sim2\%$ error due to the relative small number of points considered in the fit. }
 \label{fig_p}
 \end{center}
 }
\end{figure}

Level Spectroscopy (LS) provides an additional independent check of the RE results\cite{bosonization2}. We estimated the TLL parameter by considering the well-known relation:
\begin{equation}
K=v\pi\mathcal{C}
\end{equation} 
where $\mathcal{C}$ is the compressibility of the system. We employed exact diagonalization with PBC up to $L=24$ system sizes to estimate $v$ from the CFT relation:
\begin{equation}
\epsilon_{gs}(L)=\epsilon_0+\frac{v c\pi}{6L^2}+...
\end{equation}
where $\epsilon_{gs}(L)$ is the energy density of the ground state at size $L$, $\epsilon_0$ being the thermodynamic value. The compressibility can then be calculated as:
\begin{equation}
\mathcal{C}=L(E_L(N-1)+E_L(N+1)-2E_L(N))
\end{equation}
where $E_L(M)$ denotes the ground state energy of a system of $L$ sites with $M$ particles. Typical scalings of both $\epsilon_{gs}$ and $K$ are presented in Fig.\ref{fig_LS}. LS displays a systematic overestimation error of order $\sim5\%$ due to finite-size effects (see Fig. \ref{fig_p}), effect that we verified for some sample points by calculating $\mathcal{C}$ using a multi-target DMRG method \cite{sp_ortolani} up to $L=44$.

%\tableofcontents

%\section{Introduction}


\begin{thebibliography}{39}
\bibitem{note1}For $\bar{n}=1/p$, logarithmic corrections emerge around $K=2/(p^2)$, whose influence on the SF-CDW PT is, in our case, negligible.
\bibitem{note2}Small quantitative changes may arise when considering the full dipolar interaction, but they are not crucial here.
\end{thebibliography}
\end{document}